\documentclass{article}
\usepackage{spconf,amsmath,amssymb,graphicx}

\usepackage{enumitem}
\usepackage{stfloats}
\setlist{nosep, leftmargin=14pt}

\usepackage{mwe} % to get dummy images
\usepackage{algorithm}
\usepackage{color}
\usepackage{changes}
\usepackage[noend]{algpseudocode}
\title{Unsupervised Anomaly Detection in 3D Brain MRI using Deep Learning with impured training data}
\name{
\begin{tabular}{lll}
Finn Behrendt*$^{\star }$  &
Marcel Bengs*$^{\star }$&
Frederik Rogge$^{\star }$ \\ 
Julia Krüger$^{\dagger}$ &
Roland Opfer$^{\dagger}$&
Alexander Schlaefer$^{\star}$
\end{tabular}\thanks{ * Both authors contributed equally.}}

\address{$^{\star}$ Institute of Medical Technology and Intelligent Systems,\\ Hamburg University of Technology,  Hamburg, Germany \\  $^{\dagger}$Jung Diagnostics GmbH, Hamburg, Germany}

\begin{document}

%\ninept
%

\maketitle

\begin{abstract} %150-250 words
% TBD: mehr auf clinical data rraity eingehen -> motiviert nutzen von nicht so gut gelabelten sets o.Ä.
%Unsupervised anomaly detection (UAD) methods have shown promising results for this task, without the need for training sets where all anomalies occur in equal frequency. A remaining precondition for UAD methods is a training distribution that solely contains healthy samples. However, the effect of overseen unhealthy samples in the training distribution of UAD methods remains unexplored for Brain MRI.
The detection of lesions in magnetic resonance imaging (MRI)-scans of human brains remains challenging, time-consuming and error-prone. Recently, unsupervised anomaly detection (UAD) methods have shown promising results for this task. These methods rely on training data sets that solely contain healthy samples. Compared to supervised approaches, this significantly reduces the need for an extensive amount of labeled training data. However, data labelling remains error-prone. We study how unhealthy samples within the training data affect anomaly detection performance for brain MRI-scans. For our evaluations, we consider three publicly available data sets and use autoencoders (AE) as a well-established baseline method for UAD. We systematically evaluate the effect of impured training data by injecting different quantities of unhealthy samples to our training set of healthy samples from T1-weighted MRI-scans. We evaluate a method to identify falsely labeled samples directly during training based on the reconstruction error of the AE.  Our results show that training with impured data decreases the UAD performance notably even with few falsely labeled samples. By performing outlier removal directly during training based on the reconstruction-loss, we demonstrate that falsely labeled data can be detected and removed to mitigate the effect of falsely labeled data. Overall, we highlight the importance of clean data sets for UAD in brain MRI and demonstrate an approach for detecting falsely labeled data directly during training.
\end{abstract}

\section{Introduction}
% Why Brain MRI?
Brain MRI is commonly used for the diagnosis and treatment of neurological diseases. MRI-scans allow for the detection and delineation of brain abnormalities without the need for harmful radiation. In clinical practice the MRI-scans are manually assessed by physicians which is time-consuming and error-prone, especially for unexpected anomalies \cite{Drew.2013,Kim.2014}. \\
% Hence, a support tool for automatic detection of lesions would have the potential to reduce the workload for physicians, save costs and improve the diagnostic procedure. 
% Why unsupervised?
Supervised deep learning methods have shown promising results for the detection and segmentation of anomalies in brain MRI-scans \cite{Lundervold.2019}, relying on large-scale annotated data sets. 
However, the requirement of large-scale data sets with pixel-level annotations represents a hurdle since they are costly and time-consuming to obtain. 
Furthermore, the flexibility to detect rare diseases is limited as training of supervised methods is challenging in the case of unbalanced data \cite{Johnson.2019}. 
% UAD
Deep learning for unsupervised anomaly detection follows the concept of identifying abnormalities by learning from a reference data set of healthy data. Different deep learning models have been used within the UAD framework, mainly focusing on autoencoder-based approaches \cite{Baur.2019AE,Atlason.2019}, generative adversarial networks \cite{Schlegl.2019} or the combination of both \cite{baur2021autoencoders}. Typically, a convolutional neural network (CNN) learns to compress and reconstruct the anatomical features of healthy brains but fails to reconstruct unseen anomalies. Afterward, detection can be performed by considering the reconstruction errors. Thus, in theory, UAD methods can detect arbitrary anomalies but require a large-scale training data set that solely contains healthy samples. These large-scale training data sets could be obtained from routine MRI-scans. However, data labelling is affected by human errors \cite{Kim.2014}, leading to impured training data sets. In contrast to that, a clean data set is a precondition for many UAD methods. Learning with impured data sets forces UAD methods to learn to compress and reconstruct also unhealthy samples which conflicts with the fundamental concept of UAD methods.\\ In this work, we study how impured training data affects UAD performance in brain MRI-scans and evaluate a first approach to counteract this problem directly during training. 
We use 3D autoencoders (AE) as a well-established baseline using the publicly available brain MRI data sets IXI and OASIS-3 as clean data sets for our study.
We systematically evaluate UAD performance on different data set compositions, containing different amounts of unhealthy samples.  To this end, we add different proportions ranging from 3\% up to 12\% of brain MRI-scans that contain unhealthy brains to our clean training set that contains solely healthy labeled MRI-scans and evaluate UAD performance afterwards. As a first step towards outlier removal from the impured training data, we identify and remove outliers based on the loss function of the AE directly during training. 
\section{Methods}
%% Architecture (small)
% Data
\subsection{Data Set}
\begin{figure*}[t]
\centering
\includegraphics[width=\textwidth]{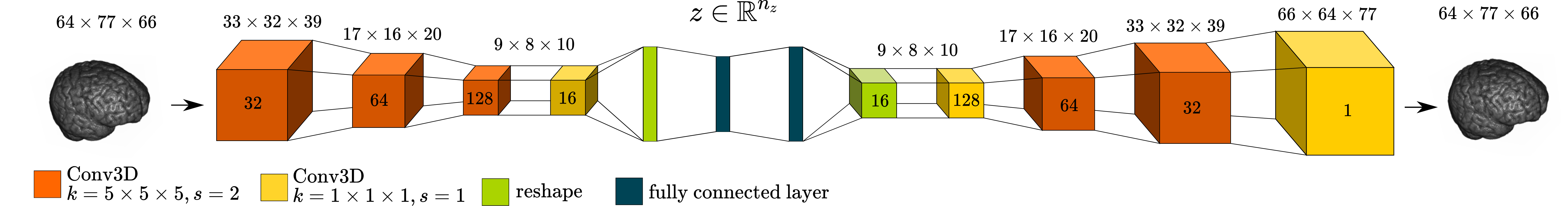}
\caption{Our 3D-AE architecture. An input volume $x$ is encoded to the latent space $z$ and the output volume is reconstructed from $z$. Numbers above each convolutional block denote the spatial input size and numbers within each convolutional block denote the number of feature maps.}
\label{fig:model}
\end{figure*}
% MixedNormals
%For training, we use a data set consisting of 600 T1-weighted MRI-scans. The MRI-scans are acquired by different scanner models from 22 different vendors. While the axial resolution varies from X to Y, the majority (N=ZZZ) of the scans is acquired with a resolution of \SI{1}{mm}. The slice thickness ranges from XX mm to YY mm. The majority (N=XXX) of the scans are acquired with a slice thickness of ZZ mm. X, Y, Z MRI-scans are acquired with a field strength of 1 T, 1.5 T and 3 T respectively. The mean age of the scanned patients is XX years with a standard deviation of YY years. The data set was acquired during clinical routine with standard 3D gradient echo sequences. The MRI-scans were analized and assessed as healthy by jung diagnostics GmbH.
% BraTS
For training, we use two publicly available data sets and combine 496 T1-weighted healthy samples from the OASIS-3 data set \cite{LaMontagne.2018} with 577 T1-weighted healthy samples from the IXI dataset\footnote{https://brain-development.org/ixi-
dataset/}. This healthy training set is denoted as Train$_{clean}$. For evaluation, we use T1-weighted MRI-scans from the publicly available Multi-modal Brain Tumor Segmentation Challenge 2019 (BraTS19)  \cite{Menze.2015,bakas2017advancing} data set. The BraTS19 data set contains 335 MRI-scans of subjects with high- and low-grade gliomas. We follow common preprocessing strategies \cite{baur2021autoencoders} by applying skull stripping, denoising, and normalization to all samples. We ensure an isotropic resolution of $1 \times 1 \times 1$ voxels by interpolation, crop the empty background, and apply zero-padding to all samples to achieve a consistent resolution of $160 \times 192 \times 166$ voxels. The samples are then resized by a factor of 0.4 in each dimension to a resolution of $64 \times 77 \times 66$ voxels for computational efficiency.  To obtain an impured training data set, we include a varying number of unhealthy samples from the BraTS19 data set that are not used for evaluation to our training set Train$_{clean}$. By including approximately 3\%, 6\% and 12\%  of unhealthy samples to our training set, we obtain training sets with different levels of impurity, denoted as Train$_{3\%}$, Train$_{6\%}$ and Train$_{12\%}$ respectively. We randomly split the healthy data set into a healthy training set containing 600 samples and a healthy validation set of 153 samples for hyperparameter tuning. For evaluation, we use 160 samples from the BraTS19 data set together with 320 healthy test samples. Note that the 160 samples for evaluation are independent and not used to impure our training data set.
%The Brats and Stroke data sets are split into an abnormal validation set and an abnormal test set, consisting of XX and YYY samples respectively. 
% Stroke
% Data sets and explain outlier injection 
\subsection{Deep Learning Methods}
% \begin{algorithm}[!hb]
% \caption{Outlier removal algorithm. $\theta$ denotes the model parameters of the AE model $f_{\theta}$. In our experiments, we train for $m=500$ epochs, remove outlier every $n=5$ epochs and reinitialize the model parameters $\theta$ in the epochs $r\in [50,100]$. Note that batching is omitted and the batch size is set the size of the training data set here for simplicity.}\label{alg:removal}
% \begin{algorithmic}[1]
% \Require $\gamma$ \Comment{weigthing factor}
% \Require $\theta$ \Comment{initialize model parameters}
% \Require $X$ \Comment{training set}
% \Require $m$ \Comment{maximal number of training epochs}
% \Require $n$ \Comment{remove every n epochs}
% \Require $r$ \Comment{epochs for reinitialization}
% \For {$epoch$ in $[1,2,\ldots,m]$}
% % \ForAll{$x$ in $\textbf{X}$}
% \State $\hat{X} \gets f_{\theta}(X)$  \Comment{reconstruct samples}
% \State $loss \gets \mathcal{L}_{AE}(X,\hat{X})$

% % \EndFor
% \If {$epoch$ in $[n,2n,\ldots,m-n]$}  \Comment{every n epochs}
% \State $\mathcal{T}_{remove} \gets \gamma \cdot loss$ \Comment{threshold}
% \ForAll{$x$ in $X$}
% % \State $\hat{x} \gets f(\theta,x)$  \Comment{reconstruct samples}
% \State error $\gets \lvert x-f_{\theta}(x)\rvert$
% \If {$error > \mathcal{T}_{remove}$}
% \State remove $x$ from $X$
% \EndIf
% \EndFor
% \EndIf
% \State $\theta \gets  \theta - \nabla_{\theta} \cdot loss$ \Comment{update parameters}
% \If{$epoch$  in $r$}
% \State reset $\theta$  \Comment{reinitialization}
% \EndIf

% \EndFor
% \Return $\theta$
% \end{algorithmic}
% \end{algorithm}
We consider AEs for the task of UAD as they are a well-established baseline for UAD in brain MRI-scans \cite{baur2021autoencoders}. Since we study sample-wise anomaly detection, we extend the AE architecture from \cite{baur2021autoencoders} to 3D by the use of 3D-convolutions and adjust the latent space size to n$_z = 512$. This aims to make use of the spatial context of the volumetric MRI-scans as it is shown in \cite{bengs2021three}. Our network architecture is shown in Figure \ref{fig:model}. We follow the standard concept of AEs where the loss function is formulated as \begin{equation}
\mathcal{L}_{AE}=L_{Rec}(x,\hat{x}) = \frac{1}{N}\sum_{k=1}^{N} |x_{k}-\hat{x}_{k}|
\label{eq:AE_train}
\end{equation} 
with $N$ for the number of samples in the training set, $x \in \mathbb{R}^{h \times w \times d}$ and $\hat{x} \in \mathbb{R}^{h \times w \times d}$. $h,w,d$ are the height, width and depth of the volumes. 
Having trained the AE, the reconstruction error between input and reconstruction can be used as an anomaly score. For samples that are similar to the training data set, the reconstruction error is assumed to be small, compared to samples that differ from the training data set. Therefore, if the AE is trained with only healthy samples, $\mathcal{L}_{AE}$ can be used to detect unhealthy samples as outlier from the training set, based on high reconstruction errors \cite{baur2021autoencoders}. \\ Several approaches have been proposed to identify out of distribution samples in anomaly detection scenarios \cite{xia2015learning, hendrycks2018deep, 9577693}.
In this work, we adapt the general concept of Xia et al. \cite{xia2015learning} to perform outlier removal directly during training.
For images of the natural image domain, Xia et al. \cite{xia2015learning} show that $\mathcal{L}_{AE}$ can be used to discriminate and cluster between images of different classes directly during training. We adapt this method and investigate whether it is feasible to identify and remove unhealthy samples from MRI-scans of human brains within an impured training data set. Our hypothesis is that for few unhealthy samples, the reconstruction error is high compared to healthy samples already during training. Hence, we perform outlier removal based on $\mathcal{L}_{AE}$ to discriminate unhealthy samples from healthy ones during training. Therefore, we modify the training of the AE and remove samples from the training set every 5 epochs. Samples are removed based on an adaptive threshold, above which the samples are identified as outliers. The adaptive threshold is determined by 
\begin{equation}
    \mathcal{T}_{remove}^i=\gamma \cdot \mathcal{L}_{AE}^i,
\end{equation} where $\mathcal{L}_{AE}^i$ denotes the current loss of the entire training set in the current epoch $i$ and $\gamma$ denotes a weighting factor which determines the boundary between outlier and normal sample during training. We reinitialize the network's parameter after 50 and 100 epochs. This aims to reduce the effect of outliers on the final model while removing outliers from the training data set. The number of epochs have been determined, based on prior experiments using the validation set. After the last re-initialization, we train the network for another 400 epochs, again by removing outliers every 5 epochs.
%By removing outliers during training, the AE automatically cleans its training distribution which improves the ability to identify unhealthy samples as outliers after training. 
% VAE ist Praktisch dafür, da er KLD mitliefert!
% VAE / 3D VAE
\begin{figure*}[!ht]
\setlength{\tabcolsep}{1pt}
    \centering
    \begin{tabular}{c|c|c|c}
    \hspace{17mm}Clean Set (Train$_{clean}$) & $3\%$ injected (Train$_{3\%}$)& $6\%$ injected (Train$_{6\%}$)& $12\%$ injected (Train$_{12\%}$)
    \\
    \includegraphics[width=0.305\textwidth]{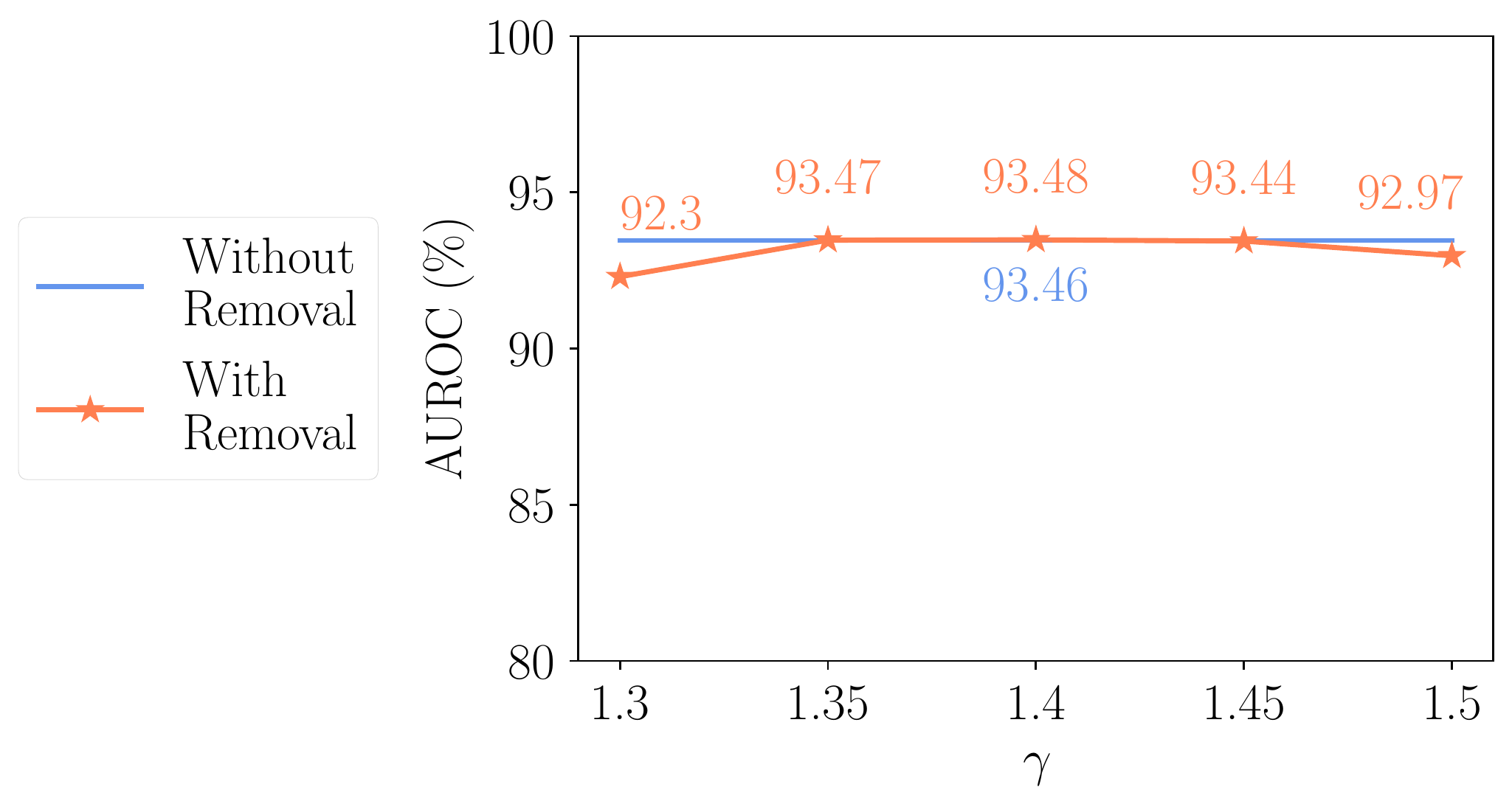} &
    \includegraphics[width=0.225\textwidth]{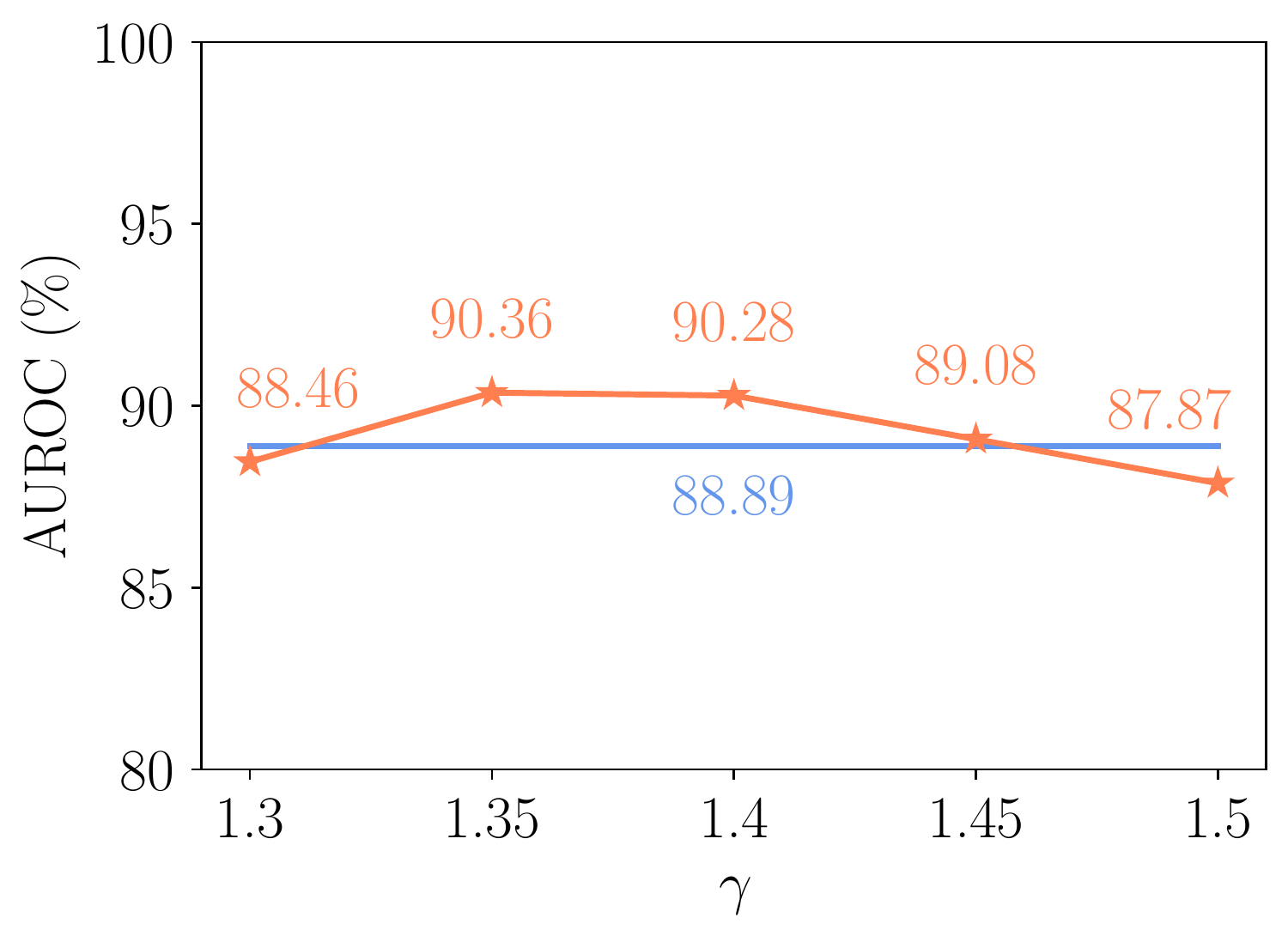} &
    \includegraphics[width=0.225\textwidth]{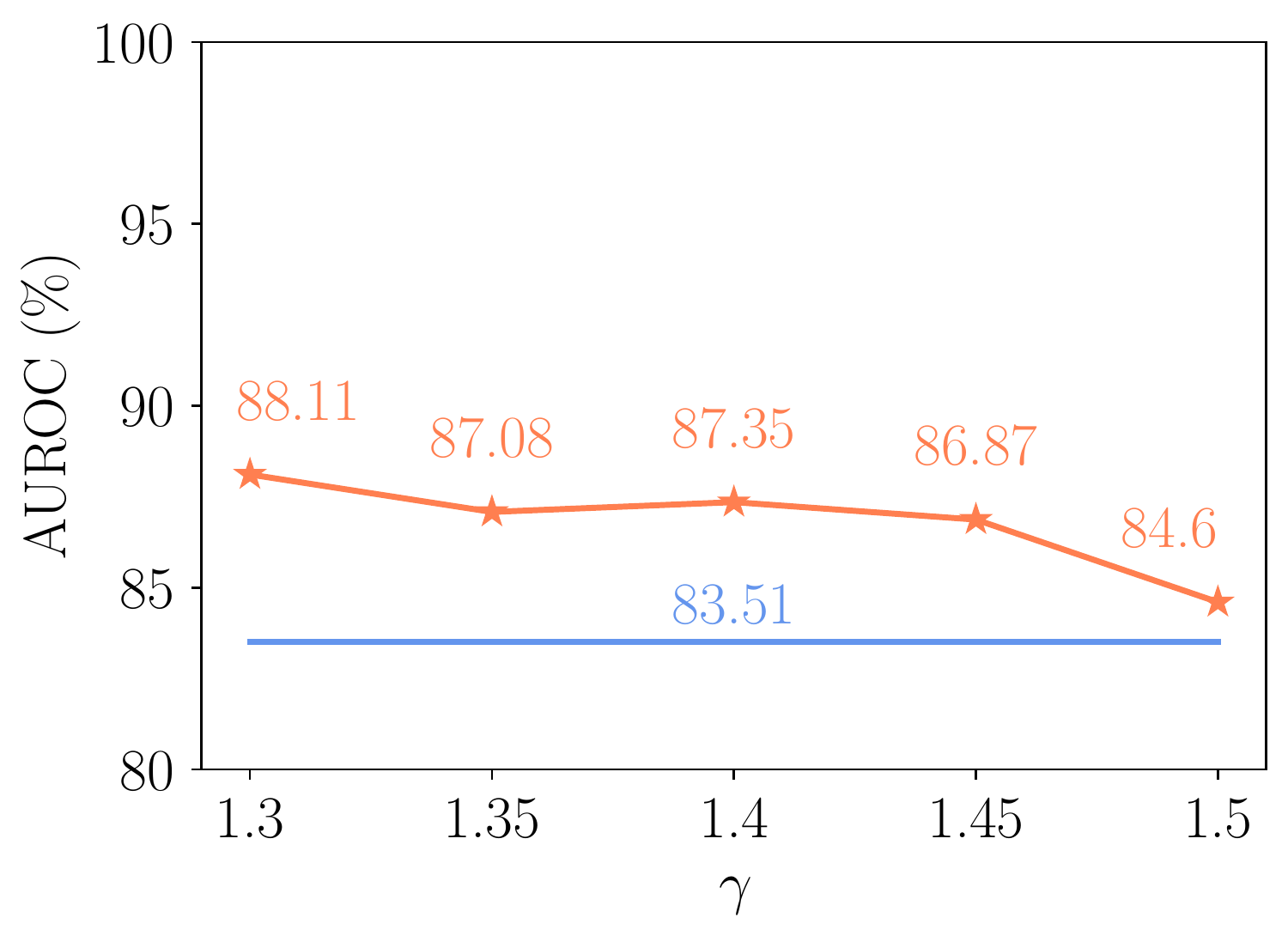}&
    \includegraphics[width=0.225\textwidth]{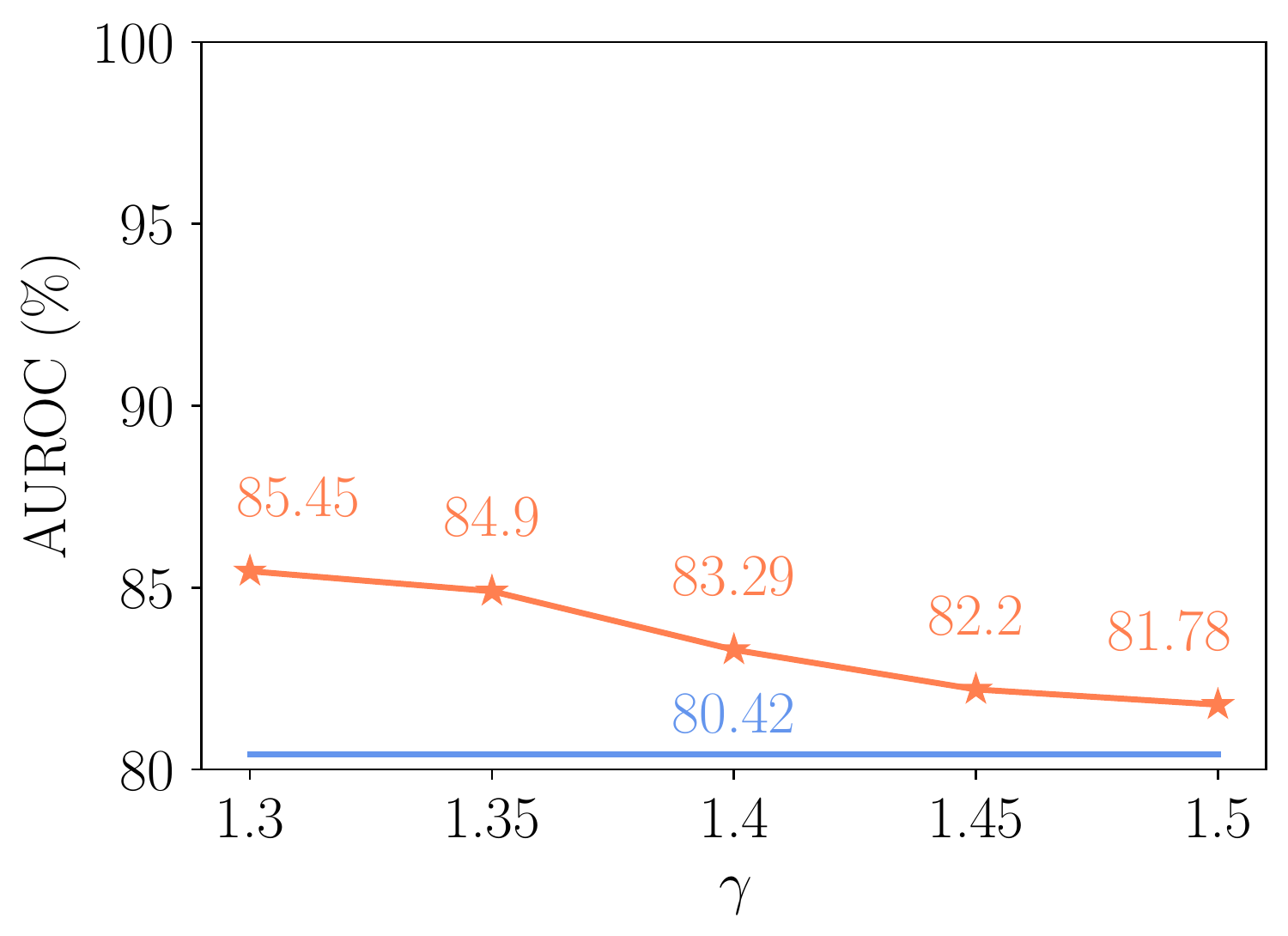} 
    \\ 
    \includegraphics[width=0.305\textwidth]{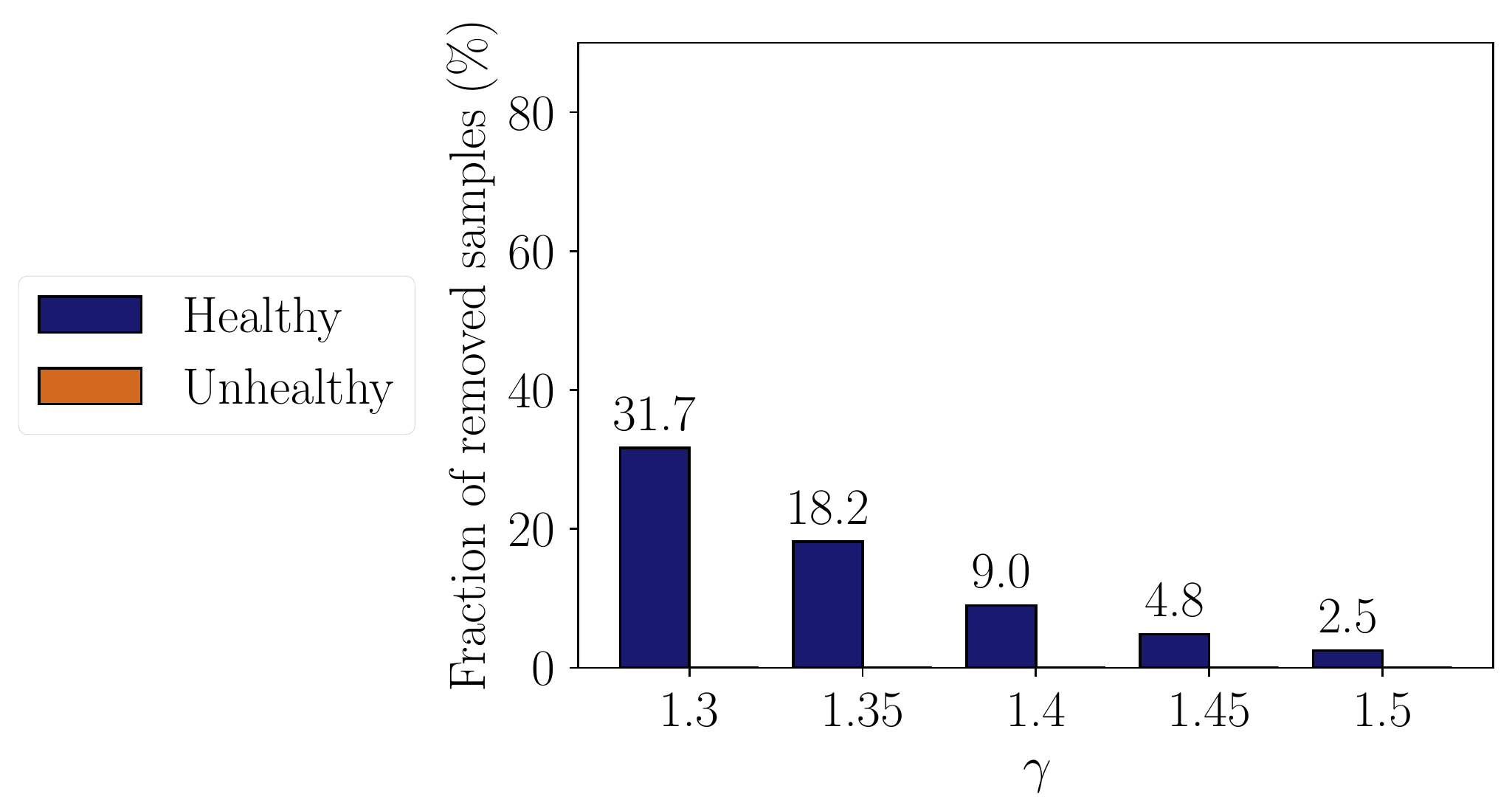} &
    \includegraphics[width=0.225\textwidth]{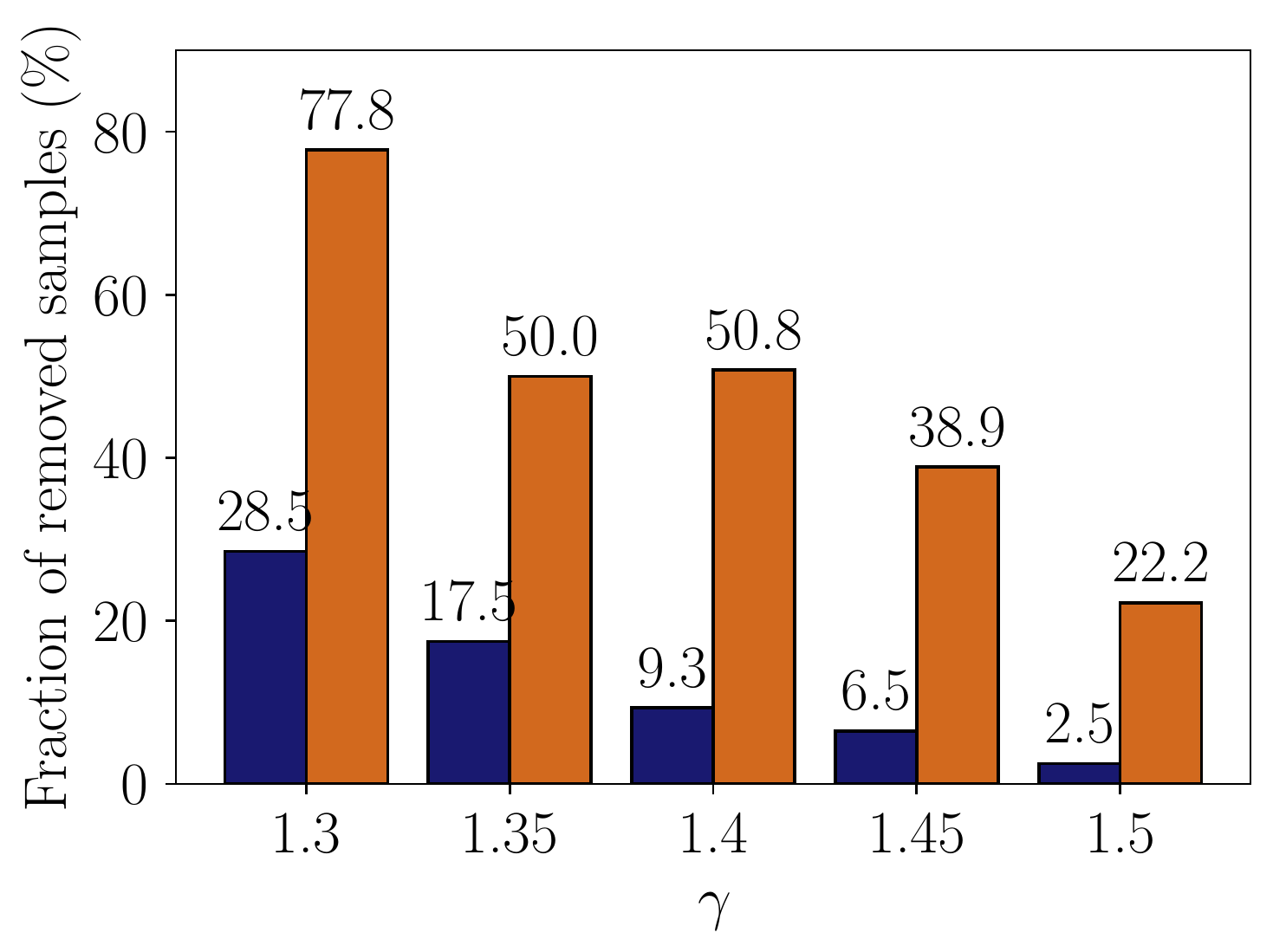} &
    \includegraphics[width=0.225\textwidth]{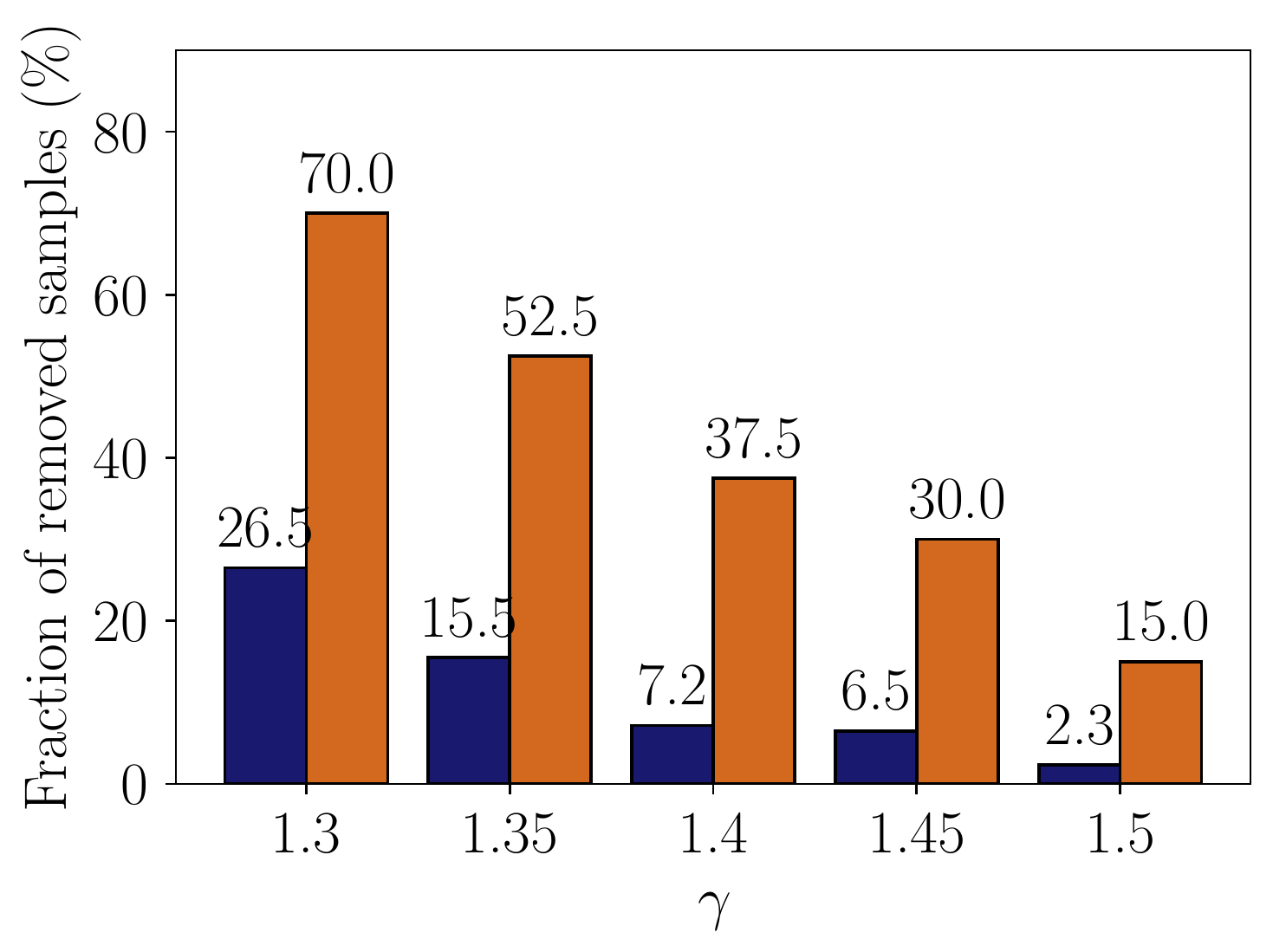} &
    \includegraphics[width=0.225\textwidth]{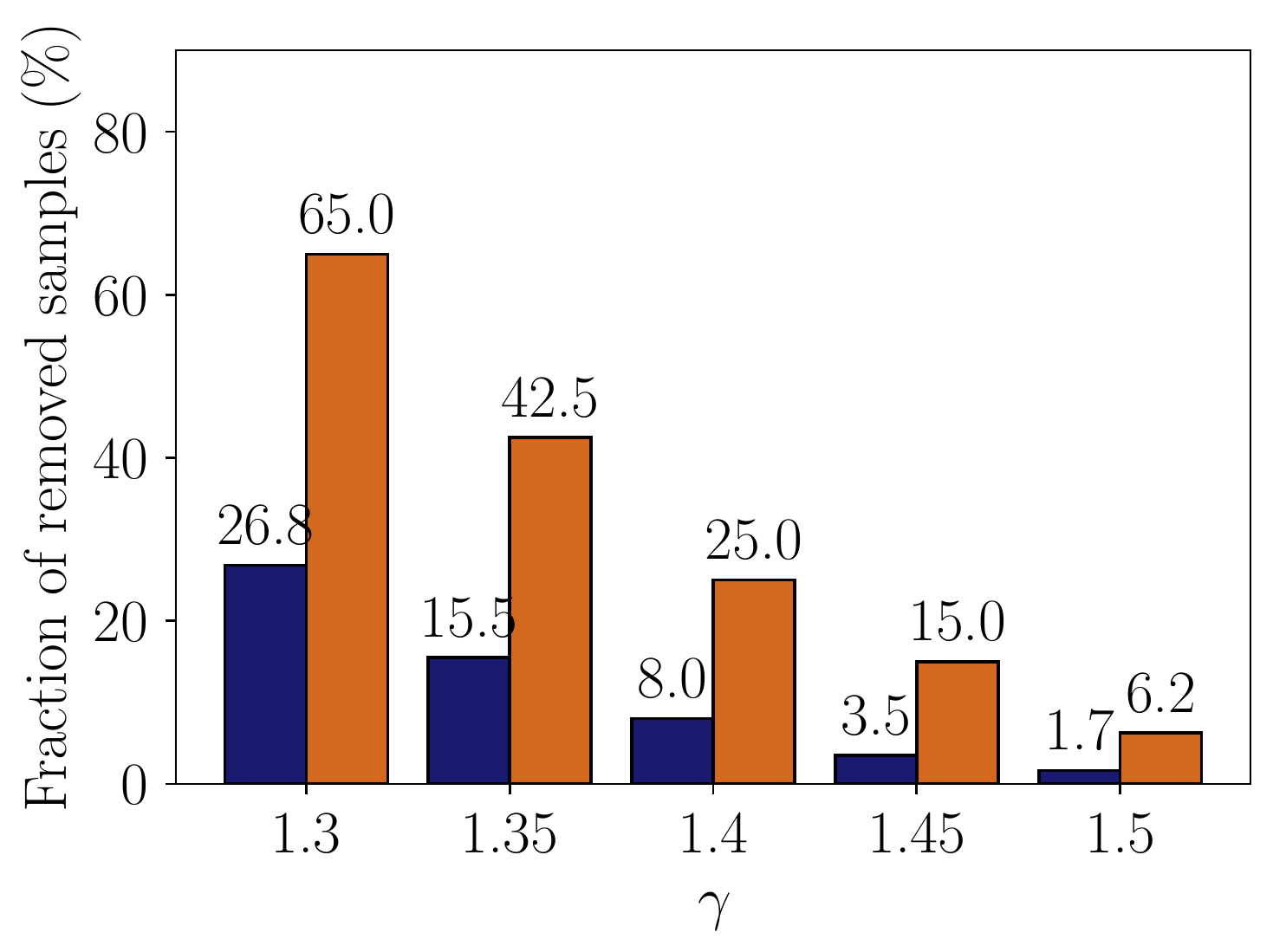} 
    \end{tabular}
   \caption{From left to right, data sets with different levels of impurity are evaluated. Top row: For different values of $\gamma$ the UAD performance regarding the AUROC with and without outlier removal is reported. Note that the baseline is independent of $\gamma$ and shows a constant AUROC. Bottom row: For different values of $\gamma$ the relative fraction of removed healthy and unhealthy samples is reported.}
    \label{tab:results}
\end{figure*}
% Outlier removal
% \subsection{Training and Evaluation}
We use Adam for optimization, a learning rate of $lr=10^{-3}$ and a batch size of $bs = 32$ for our experiments and train our models on an NVIDIA GTX 1080ti graphics card. \\
% We evaluate the detection performance between each unhealthy test set and the healthy test set by relying on the anomaly score $\mathcal{L}_{AE}$.
We use the reconstruction error between input volume and reconstructed volume as the anomaly score to discriminate between healthy and unhealthy samples during evaluation. To assess the model's performance independent of a chosen operation point, we report the AUROC. Furthermore, we report the ratio of samples removed during training relative to the number of total samples in the training data, for healthy and unhealthy samples respectively. Across our experiments, we evaluate and report $\gamma$ in the range of $\gamma \in [1.3,1.35,\ldots,1.5]$.% as this range has shown to be valid regarding the reconstruction error of the healthy validation set. 
% training and evaluation
\section{Results}
% The anomaly detection performance for different ratios of unhealthy samples in the training set and different values for the threshold weighting factor $\gamma$ is shown in Figure \ref{tab:results}. Considering the baseline AE without outlier removal, the AUROC decreases by 5.2\% to 16.2\% with increasing number of outliers. Across all experiments, outlier removal leads to increased or similar detection performance compared to the baseline without outlier removal. 
% % Looking at Train$_{clean}$ shows that outlier removal with a clean data set still leads to increasing UAD performance even though in theory no outliers are present. 
% Furthermore, Figure \ref{tab:results} provides the ratios of samples removed during training for both, normal and abnormal samples compared to the total number of healthy and unhealthy samples, for different choices of the weighting parameter $\gamma$. Naturally, smaller values for $\gamma$ show an increased number of removed samples for both, healthy and unhealthy samples. For all choices of $\gamma$ the fraction of removed samples is higher compared to the fraction of unhealthy samples. Exemplary healthy samples that are removed when performing outlier removal with Tran$_{clean}$ are shown in Figure \ref{fig:healthy_removed}.

% Man könnte das auch so abkützen:
The anomaly detection performances for different ratios of unhealthy samples in the training set and for different values of the weighting factor $\gamma$ are shown in Figure \ref{tab:results}. Considering the baseline AE without outlier removal, the AUROC decreases from 93.46 to 88.89, 83.51 and 80.42 for 3\%, 6\% and 12\% of outliers, respectively. Across all experiments, outlier removal during training leads to an increased or similar detection performance across all values of $\gamma$ with notable performance improvements for Train$_{6\%}$ and Train$_{12\%}$. For all choices of $\gamma$, the fraction of removed healthy samples is higher compared to the fraction of removed unhealthy samples. Overall, smaller values for $\gamma$ increase the number of removed samples for both, healthy and unhealthy samples.
Exemplary healthy samples that are removed when performing outlier removal with Train$_{clean}$ are shown in Figure \ref{fig:healthy_removed}. \\ 
Figure \ref{fig:grids} shows exemplary reconstructions of AE models, trained with different ratios of unhealthy samples in the training set. With an increasing amount of unhealthy samples in the training set, the AE learns to reconstruct not only healthy brains but also anomalies from the unhealthy samples. When removing outliers during training, fewer anomalies are reconstructed and the overall reconstruction error of unhealthy brains is increased.  
% In our experiments, $\gamma=1.4$ has shown the best results based on the healthy validation set.  
%Figure \textbf{Fig} shows example brain MRI-scans of healthy samples that are falsely removed during training. %TBD: samples einfügen und gucken, ob die normal aussehen 
%Overall, the outlier removal method works best for a larger number of added outliers. For Train$_{10\%$ removing outliers during trianing improves the AUROC by 4.2\% and 1.27\% for the BraTS19 and ATLAS data set respectively. In comparison, for Train$_{20\%}$ removing outliers improves the AUROC by 4.4\% and 1.8\% for the Brats19 and ATLAS data set respectively. 
%% What Plots?

% \begin{figure}[t]
%     \centering
%     \includegraphics[width=0.50\textwidth]{AUC_removal.pdf}
%   \caption{Comparison of UAD performance regarding the AUROC with and without outlier removal for different proportions of unhealhty data in the training distribution. Metrics are reported in percent.}
%     \label{tab:results}
% \end{figure}
%\begin{figure}[t]

%    \centering
%    \includegraphics[width=0.50\textwidth]{barplot_fraction_removed_samples_80.pdf}
%   \caption{Fraction of removed samples relative to their total amount within Train$_{20\%}$ for different thresholds $\gamma$. Large values of $\gamma$ indicate a large threshold above which samples are removed and vice-versa.}
%    \label{fig:outlier_ratio}
%\end{figure}
% show the effect of outlier removal in comparison to the baseline for different injections
% example grids of reconstructions and reco errors
\section{Discussion}
\begin{figure}[ht]
    \centering
    \begin{tabular}{c}
     No injection (Train$_{clean}$) without removal\\
    \includegraphics[width=0.45\textwidth]{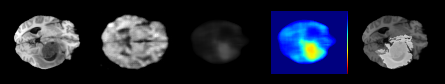} \\
     No injection (Train$_{clean}$) with removal\\
    \includegraphics[width=0.45\textwidth]{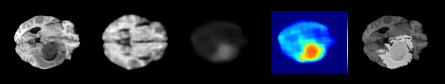}\\
    %  3\% injected (Train$_{3\%}$) without removal\\
    % \includegraphics[width=0.45\textwidth]{BraTS19_CBICA_AAB_1_t1.ni_30_Grid_18.png} \\
    %  3\% injected (Train$_{3\%}$) with removal\\
    % \includegraphics[width=0.45\textwidth]{BraTS19_CBICA_AAB_1_t1.ni_30_Grid_40.png}\\
    %  6\% injected (Train$_{6\%}$) without removal\\
    % \includegraphics[width=0.45\textwidth]{BraTS19_CBICA_AAB_1_t1.ni_30_Grid_inject40.png}\\
    %  6\% injected (Train$_{6\%}$) with removal\\
    % \includegraphics[width=0.45\textwidth]{BraTS19_CBICA_AAB_1_t1.ni_30_Grid_40_removed.png}\\
    12\% injected (Train$_{12\%}$) without removal\\
    \includegraphics[width=0.45\textwidth]{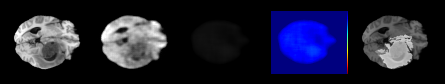}\\
     12\% injected (Train$_{12\%}$) with removal\\
    \includegraphics[width=0.45\textwidth]{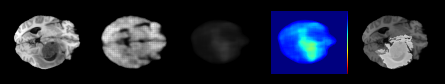}
    \end{tabular}
   \caption{Exemplary reconstruction of unhealthy samples for different levels of impurity with and without outlier removal for $\gamma = 1.3$. From left to right, input, reconstruction, reconstruction error, a heatmap of the reconstruction error and the ground truth are shown. From top to bottom, experiments for Train$_{clean}$ and Train$_{12\%}$ are shown without and with outlier removal, respectively.}
    \label{fig:grids}
\end{figure}

\begin{figure}[!ht]
    \centering
    \includegraphics[width=0.2\textwidth]{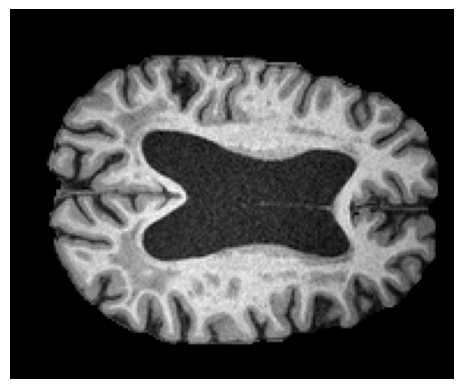} 
        \includegraphics[width=0.2\textwidth]{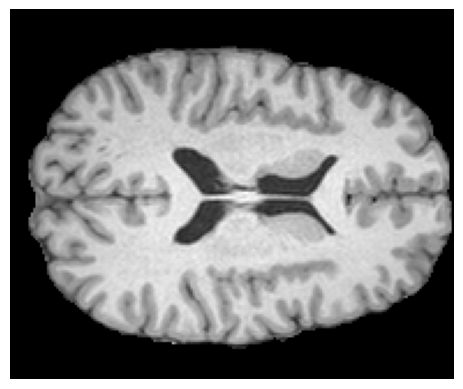}
        
   \caption{Exemplary slices of two healthy MRI-scans that are labeled as healthy and removed by our algorithm. Left: MRI-scan from the IXI data set. Right: MRI-scan from the OASIS-3 data set}
    \label{fig:healthy_removed}
\end{figure}

Clean training data sets consisting of only healthy samples is a precondition for UAD with deep learning that is rarely questioned. However, for large-scale data sets, this precondition is hard to fulfill since overseen anomalies and human errors are likely within the labelling process \cite{Drew.2013,Kim.2014}.
%We study the influence of unhealthy samples in the healthy training data and evaluate an approach for outlier removal directly during training.  
Our results show that training with unhealthy samples decreases the UAD performance notably. This is not surprising considering the concept of UAD but what is striking is that even with a small quantity of 3\% of unhealthy samples in the training data the AUROC decreases from 93.46\% to 88.89\%, see Figure \ref{tab:results}. This demonstrates that the AE rapidly adapts to overseen anomalies during training, even in small quantities as it can be seen in Figure \ref{fig:grids}. Hence, even though less labelling effort is required for UAD, our findings highlight that precise data labelling is vital. This indicates that while deep learning for UAD can detect any anomaly in theory it is limited to those that are strictly not present during training. Hence, detecting arbitrary anomalies that are not considered during data labelling by radiologists seems difficult when training data is not questioned. \\ As a first step towards analyzing the training data, we investigate the feasibility of removing outliers directly during training. Our results in Figure \ref{tab:results} demonstrate that unhealthy samples can be detected already during the training process. By removing the samples identified as outliers, the anomaly detection performance can be improved notably, especially for training data with larger quantities of unhealthy samples. Outlier removal during training reduces the adaptation of the AE to the unhealthy samples in the training set (see Figure \ref{fig:grids}) and thus improves the ability to detect unhealthy samples in the evaluation.
As it is shown in Figure \ref{tab:results}, smaller values of $\gamma$ result in larger fractions of removed healthy and unhealthy samples. While for Train$_{clean}$ and Train$_{3\%}$ no notable difference can be observed in the AUROC for varying values of $\gamma$, for Train$_{6\%}$ and Train$_{12\%}$, smaller values of $\gamma$ improve the AUROC. In general, $\gamma$ controls the number of removed samples that have to be re-labeled. While our results demonstrate similar or improved performance for all values of $\gamma$, in practice the level of impurity of a given training data set is unknown and $\gamma$ can be considered a hyperparameter, where different values can be evaluated easily. 
% Therefore, for higher levels of impurity in the training data, a smaller value for $\gamma$ is beneficial. Hence, for very trustful data sets, where less unhealthy samples are expected, larger values for $\gamma$ are preferable. Vice versa, for training data sets where a larger amount of unhealthy samples are suspected, smaller values of $\gamma$ help to identify the unhealthy samples as outliers. Thus, $\gamma$ can be seen as quantification of how trustful the training data set is and controls the number of removed samples that have to be re-labeled.
% It stands out that even for the data set that only contains healthy labeled samples the performance is improved with outlier removal.
In our experiments, healthy samples from the clean data set that are removed during training show a varying appearance of brains. These removed samples are not necessarily unhealthy especially for older patients (see Figure \ref{fig:healthy_removed}). Thus, they contain information, e.g., of the homogeneity and possible imbalances of the training data set regarding the age distribution or scanners and could be inspected by experts to post-screen and maintain UAD training data sets. Hereby, as it is stated above the weighting factor $\gamma$ would determine the sensitivity of the outlier removal. Hence, applying our evaluated outlier removal approach could be used as an effective pre-processing step of UAD training data sets. While differences in the age distribution can effect outlier removal, age information could be included in the training process similar to \cite{2022arXiv220113081B}. Our approach could be applied to more anomaly detection scenarios and diseases. However, a large-scale public benchmark anomaly detection data set of various disease types which would help our work and the field of UAD in general is missing \cite{baur2021autoencoders}. 
\section{Conclusion}
We study the effect of impured training data sets on UAD performance. Our results demonstrate that even few falsely labeled samples have a critical impact. To counteract, we evaluate an outlier removal approach directly during training. Our findings show that falsely labeled samples can be detected and removed during training, which mitigates the effect of impured training data sets and helps to revise UAD training data sets regarding domain shifts and imbalances. Overall, we demonstrate the significance of clean data sets for UAD, which should be considered in future works. Also the effect of impured data to other UAD methods like Adversarial autoencoders or GANs could be elaborated together with outlier removal approaches.
% Moreover, further work could emphasis on other methods for outlier detection, also for other UAD methods.
\\ \\ 
\textbf{Ethical approval:} This work relies on publicly available OASIS-3, IXI and BraTS19 data sets. For use of this data sets, no ethics statements are necessary. 
\\ \\ 
\textbf{Funding:} This work was partially funded by Grant Number ZF4026303TS9 

\bibliographystyle{IEEEbib} 
\bibliography{strings,refs}

\end{document}